\documentclass[aps,prl,showpacs,groupedaddress,twocolumn]{revtex4}
\usepackage{graphicx}
\usepackage{amsmath}
\usepackage{bm}

\begin{document}

\title{Controllable Andreev retroreflection and specular Andreev reflection in a four-terminal
graphene-superconductor hybrid system}

\author {Shu-guang Cheng$^{1,2}$, Yanxia Xing$^3$, Jian Wang$^{3}$, and Qing-feng Sun$^{1,\star}$}

\address{
$^1$Institute of
Physics, Chinese Academy of Sciences, Beijing 100190, China\\
$^2$Department of Physics, Northwest University, Xi'an 710069, China\\
$^3$Department of Physics and the center of theoretical and
computational physics, The University of Hong Kong, Hong Kong,
China}

\begin{abstract}
We report the investigation of electron transport through a
four-terminal graphene-superconductor hybrid system. Due to the
quantum interference of the reflected holes from two
graphene-superconductor interfaces with phase difference $\theta$,
it is found that the specular Andreev reflection vanishes at
$\theta=0$ while the Andreev retroreflection disappears at
$\theta=\pi$. This means that the retroreflection and specular
reflection can be easily controlled and separated in this device. In
addition, due to the diffraction effect in the narrow graphene
nanoribbon, the reflected hole can exit from both graphene
terminals. As the width of nanoribbon increases, the diffraction
effect gradually disappears and the reflected hole eventually exits
from a particular graphene terminal depending on the type of Andreev
reflection.
\end{abstract}

\pacs{74.45.+c, 73.23.-b, 74.78.Na} \maketitle

\maketitle

Graphene, a single layer honeycomb lattice consisting of carbon
atoms, has attracted considerable attentions in condensed matter
community recently.\cite{ref1,ref2,ref3,ref4} The unique band
structure of graphene with a linear dispersion relation near the
Dirac-points leads to many peculiar properties, such as the
low-energy Dirac-like quasi-particle dispersion relation and the
relativistic-like behaviors.\cite{ref3,ref4} Very recently, people
begun to investigate graphene-superconductor hybrid
systems.\cite{ref5,ref6,ref7,ref8,aref1,ref9} A unique and
interesting phenomenon, the specular Andreev reflection (different
from the usual Andreev reflection), was predicted to occur at the
interface of the graphene and superconductor.\cite{ref5} It was
discovered fifty years ago,\cite{ref10} that near the interface of a
conductor and superconductor an incident electron from the metallic
side is retro-reflected as a hole and a Cooper pair is created in
the superconductor, a process known as the Andreev reflection. When
the bias is smaller than the superconductor gap, the conductance of
the metal-superconductor hybrid device is mainly determined by the
Andreev reflection. For the graphene-superconductor system, in
addition to the Andreev retroreflection, an unusual Andreev
reflection, the specular Andreev reflection may occur, in which the
direction of reflected hole is along the specular
direction.\cite{ref5} From the band structure point of view, if
electron-hole conversion is intraband: both incident electron and
reflected hole are from the same band (conduction or valence band),
this corresponds to the usual Andreev retroreflection. The specular
Andreev reflection occurs if the electron-hole conversion is
interband: the incident electron and reflected hole are,
respectively, in the conduction and valence bands. Note that in
two-terminal superconductor-graphene device, both specular
reflection and retroreflection occur. It is highly desirable to
control and separate these Andreev reflections experimentally. It is
the purpose of this letter to achieve this goal.

In this letter, we study a four-terminal graphene-superconductor
device which consists of two superconductor terminals with the phase
difference $\theta$ and two graphene terminals (see Fig.1a). By
using the non-equilibrium Green function method, the current as well
as the Andreev reflection coefficients are calculated. Our result
shows that due to the quantum interference of reflected holes from
two superconductor terminals 2 and 4 different Andreev reflection
processes can be selected by tuning the phase difference $\theta$.
When $\theta=0$ only the Andreev retroreflection occurs and the
specular Andreev reflection is prohibited while for $\theta=\pi$
only the specular Andreev reflection occurs and retroreflection
vanishes. Therefore it is very easy to control the specular Andreev
reflection and Andreev retroreflection by simply tuning the
superconductor phase difference $\theta$. In addition, the direction
of the reflected hole, which is along either the graphene terminal-1
or terminal-3 depending on the type of Andreev reflection, can only
be exhibited for large samples. When the sample size is comparable
to the wavelength of reflected hole, however, the diffraction effect
dominates so that the reflected hole can exit from both graphene
terminals.

The four-terminals device we considered consists of a zigzag edged
graphene nanoribbon sandwiched by two superconductor terminals, as
shown in the Fig.1a.\cite{anote2} In the tight-binding
representation, the Hamiltonian of the clean graphene nanoribbon is
given by\cite{ref12} $H_{G}=\sum_{i\sigma} E_0 a_{i\sigma}^\dagger
a_{i\sigma}+\sum_{<ij>\sigma} t a_{i\sigma}^\dagger a_{j\sigma}$,
where $a_{i\sigma}^\dagger$ ($a_{i\sigma}$) is the creation
(annihilation) operator at the site $i$. The on-site energy $E_0$ is
the reference energy for Dirac-point, which can be controlled
experimentally by the gate voltage. Two superconductor terminals are
represented by BCS Hamiltonian, $H_{S\alpha}=\sum_{{\bf
k}\sigma}\varepsilon_k C_{{\bf k}\sigma,\alpha}^\dagger C_{{\bf
k}\sigma,\alpha}+\sum_{\bf k} ( \Delta_{\alpha} C_{{\bf
k}\downarrow,\alpha} C_{-{\bf k}\uparrow,\alpha}+ \Delta_{\alpha}^*
C_{-{\bf k}\uparrow,\alpha}^\dagger C_{{\bf
k}\downarrow,\alpha}^\dagger)$, where $\alpha=2,4$ is the index of
the superconductor terminal and $\Delta_{\alpha} =\Delta
e^{i\theta_{\alpha}}$ with the superconductor gap $\Delta$ and phase
$\theta_{\alpha}$. The coupling between superconductor terminal
$\alpha$ and graphene is described by $H_{T\alpha}=\sum_{i\sigma} t
a_{i\sigma}^\dagger C_{\alpha,\sigma}(x_i)+h.c.$. Here $x_i$ is the
horizonal position of the carbon atom $i$ and $C_{\alpha,\sigma}(x)
=\sum_{k_x,k_y} e^{ik_x x} C_{{\bf k}\alpha,\sigma}$.\cite{ref13} So
the total Hamiltonian is $H=H_{G}+\sum_{\alpha=
2,4}(H_{S\alpha}+H_{T\alpha})$.

\begin{figure}
\includegraphics[bb=32 52 567 281, scale=0.44, clip=]{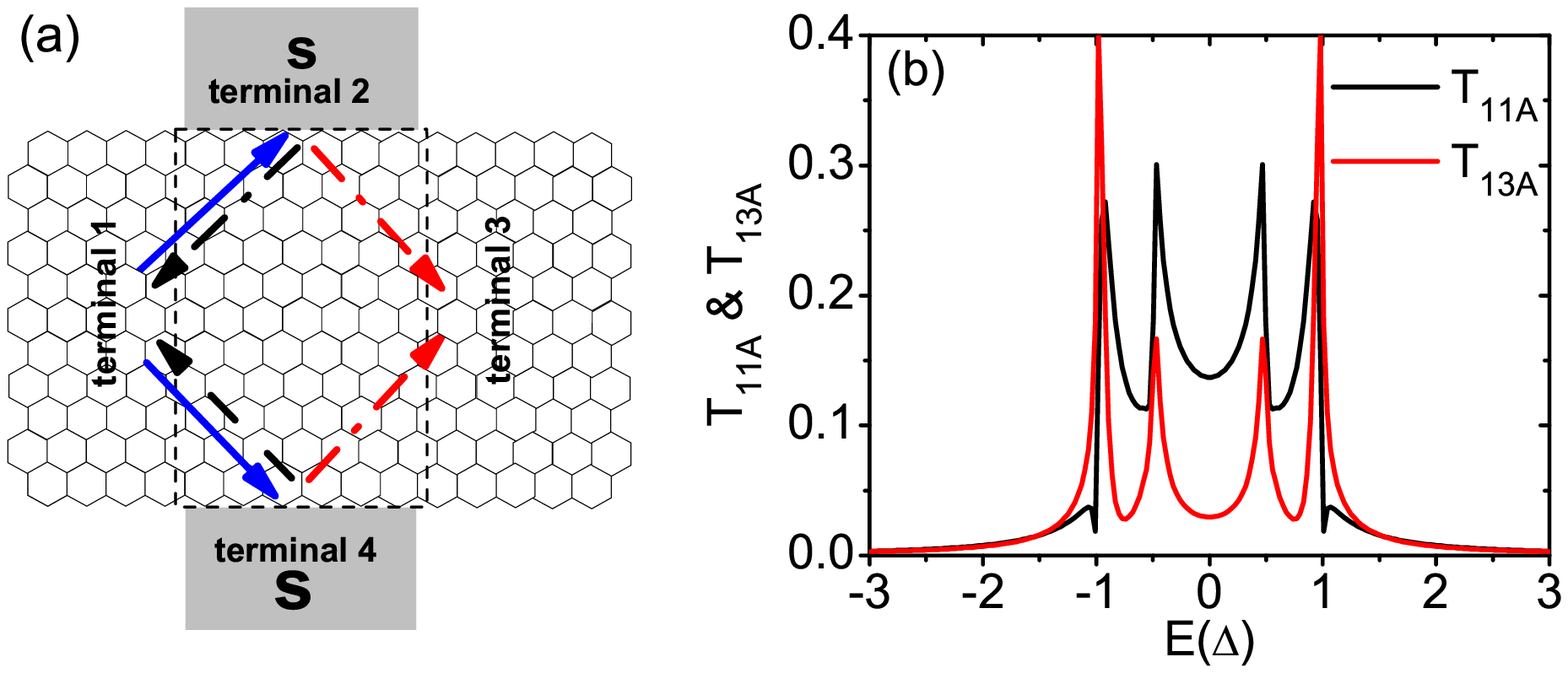}\\
\caption{(color online) (a) is the schematic diagram for
four-terminal graphene-superconductor device. In this diagram, the
width of graphene nanoribbon is $W=6$. (b) $T_{11A}$ and $T_{13A}$
vs. the energy $E$ for a three-terminal device with $E_0=-0.5\Delta$
and $W=25$.
 }
\end{figure}

Using the Heisenberg equation of motion,\cite{ref14} the current
flowing from the graphene terminal 1 to the scattering region is
found to be
\begin{eqnarray}
 I_{1}&=&\frac{2e}{\hbar} \int\frac{dE}{2\pi}
[(f_{1+}-f_2) T_{12} +(f_{1+}-f_4) T_{14}\nonumber
\\ &+&
(f_{1+}-f_{3-})T_{13A}+ (f_{1+}-f_{1-})T_{11A} \nonumber\\ &+&
(f_{1+}-f_{3+})T_{13}].
\end{eqnarray}
where $f_{\alpha \pm}(E)=1/\{\mathrm{exp}{[(E\mp
eV_{\alpha})/k_BT]}+1\}$ and
$f_2(E)=f_4(E)=1/\{\mathrm{exp}(E/k_BT)+1\}$ are the Fermi
distribution with the bias $V_{\alpha}$. Here we set the bias of two
superconductor terminals be zero ($V_2=V_4=0$). In Eq.(1),
$T_{13}(E)= \mathrm{Tr}\{{\bf\Gamma}_{1\uparrow\uparrow}{\bf
G}^r_{\uparrow\uparrow}{\bf\Gamma}_{3\uparrow\uparrow}{\bf
G}^a_{\uparrow\uparrow}\}$ and
$T_{12(14)}(E)=\mathrm{Tr}\{{\bf\Gamma}_{1\uparrow\uparrow}[{\bf
G}^r{\bf\Gamma}_{2(4)}{\bf G}_a]_{\uparrow\uparrow}\}$ are the
normal transmission coefficients from the terminal 1 to the terminal
$3$, $2$, and $4$, respectively.
$T_{11A}(E)=\mathrm{Tr}\{{\bf\Gamma}_{1\uparrow\uparrow}{\bf
G}^r_{\uparrow\downarrow}{\bf\Gamma}_{1\downarrow\downarrow}{\bf
G}^a_{\downarrow\uparrow}\}$ and
$T_{13A}(E)=\mathrm{Tr}\{{\bf\Gamma}_{1\uparrow\uparrow}{\bf
G}^r_{\uparrow\downarrow}{\bf\Gamma}_{3\downarrow\downarrow}{\bf
G}^a_{\downarrow\uparrow}\}$ are the Andreev reflection coefficients
for the incident electron coming from the terminal 1 with the hole
Andreev reflected to the terminal 1 ($T_{11A}$) or terminal 3
($T_{13A}$). Here the subscripts $\uparrow\uparrow$,
$\uparrow\downarrow$, $\downarrow\uparrow$, and
$\downarrow\downarrow$ represent the 11, 12, 21, and 22 matrix
elements in Nambu subspace. The linewidth function ${\bf
\Gamma}_{\alpha}(E)$ is defined as ${\bf\Gamma_{\alpha}}(E)=
i[{\bf\Sigma}^r_{\alpha}- ({\bf\Sigma}^r_{\alpha})^{\dagger}] $ and
$\textbf{G}^{r(a)}(E)$ are the retarded (advanced) Green functions
of central region in Nambu representation. ${\bf G}^r(E)={\bf
G}^{a\dagger}(E)=(E{\bf I-{\bf H}_{c}-\sum_{\alpha=1,2,3,4}{\bf
\Sigma^r_{\alpha}}})^{-1}$ with the Hamiltonian ${\bf H}_c$ of the
central region labeled by a rectangular area in Fig.1(a). ${\bf
\Sigma}^r_{\alpha}(E)$ is the retarded self-energy due to the
coupling to the terminal $\alpha$. $\Sigma^r_{\alpha,{ij}}(E) =t
g_{\alpha,ij}^r(E)t$, where $g_{\alpha,ij}^r(E)$ is the surface
Green function of terminal $\alpha$. For the graphene terminal 1 and
3, we have to numerically calculate their surface Green
function,\cite{ref15} while for superconductor terminal 2 and 4, the
surface Green function ${\bf g}^r_{\alpha,ij}(E)=-i\pi\rho \beta(E)
J_0[k_F(x_i-x_j)]\bigotimes\left(
                               \begin{array}{cc}
                                 1 & \Delta_{\alpha}/E \\
                                 \Delta_{\alpha}^*/E & 1 \\
                               \end{array}
                             \right)
$,\cite{ref13} where $\rho$ is the normal density of states,
$J_0[k_F(x_i-x_j)]$ is the 0-th order Bessel function with the Fermi
wave vector $k_F$, and $\beta(E)=-iE/\sqrt{\Delta^2-E^2}$ for
$|E|<\Delta$ and $\beta(E)=|E|/\sqrt{E^2-\Delta^2}$ for
$|E|>\Delta$. In numerical calculations, we set the hopping energy
$t=2.75eV$ and the length of C-C bond $a_0=0.142nm$ as in a real
graphene sample. The superconductor gap $\Delta$ is set to be
$\Delta=1meV$ and the Fermi wave-vector $k_F =1{\AA}^{-1}$.

\begin{figure}
\includegraphics[bb=30 32 584 423, scale=0.42, clip=]{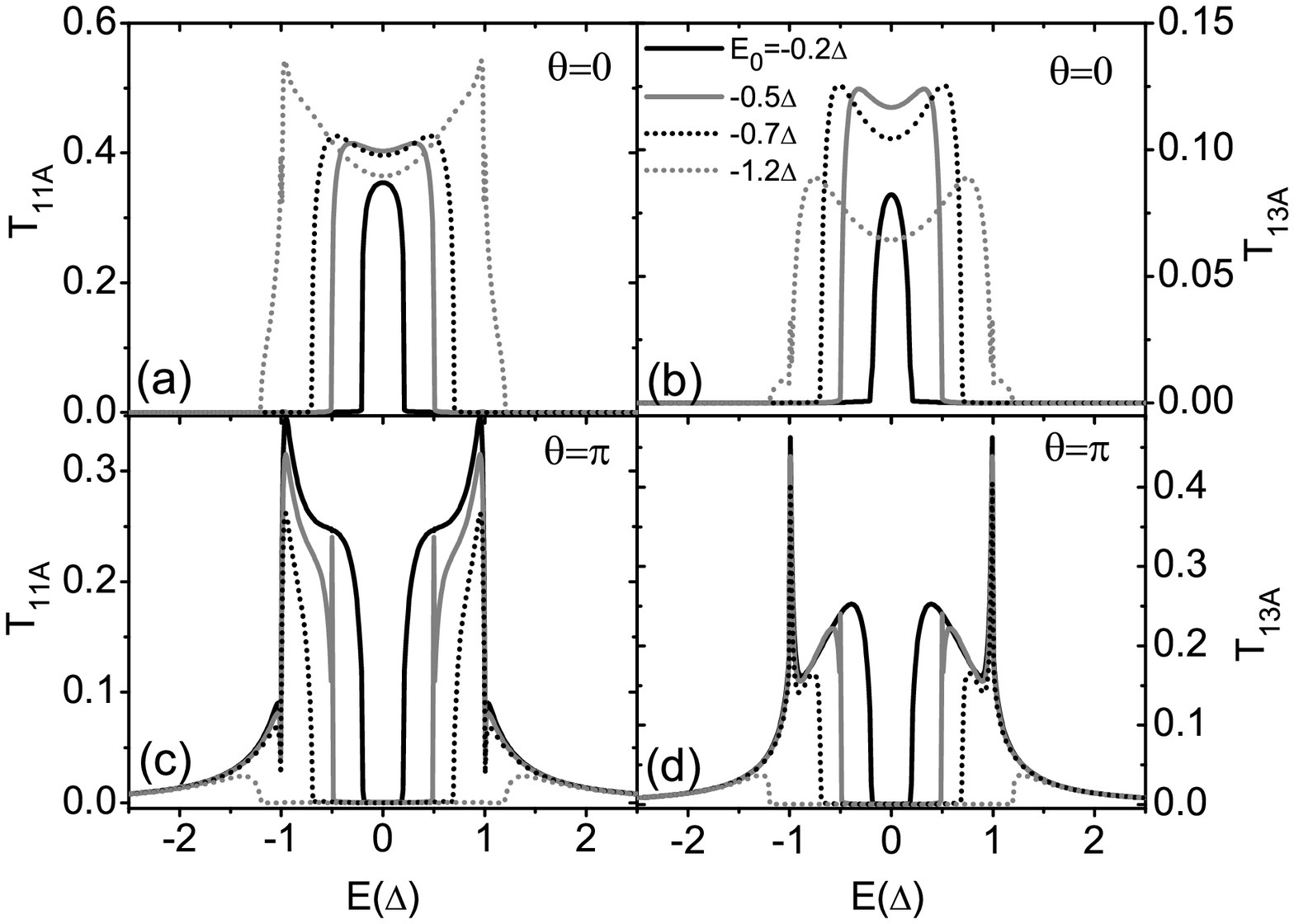}\\
\caption{$T_{11A}$ and $T_{13A}$ vs. the energy $E$ for $W=25$.}
\end{figure}

We first study a three-terminal device by decoupling one of the
superconductor terminal (2 or 4). Fig.1b shows the Andreev
reflection coefficients $T_{11A}$ and $T_{13A}$ as a function of
incident electron energies $E$. It can be seen that $T_{11A}$ and
$T_{13A}$ are quite large when the energy $E$ is within the gap
($|E|<|\Delta|$) and exhibit peaks at the Dirac points $E=\pm E_0$
and the gap edge $E=\pm \Delta$.\cite{anote1} Similar to the usual
normal-superconductor junction $T_{11A}$ and $T_{13A}$ decay quickly
when $E$ is outside of the gap.\cite{ref16}. Note that when
$|E|<|E_0|$, the incident electron and reflected hole are in the
same band (see Fig.3f) leading to the usual Andreev retroreflection.
On the other hand, for $|E|>|E_0|$, the incident electron and
reflected hole are, respectively, in the conduction and valence
bands (see Fig.3f) giving rise to the specular Andreev reflection.
The above results show that both retroreflection and specular
reflection occur with large amplitudes in the three-terminal device
(with only one superconductor terminal).

\begin{figure}
\includegraphics[bb=24 30 570 459, scale=0.42, clip=]{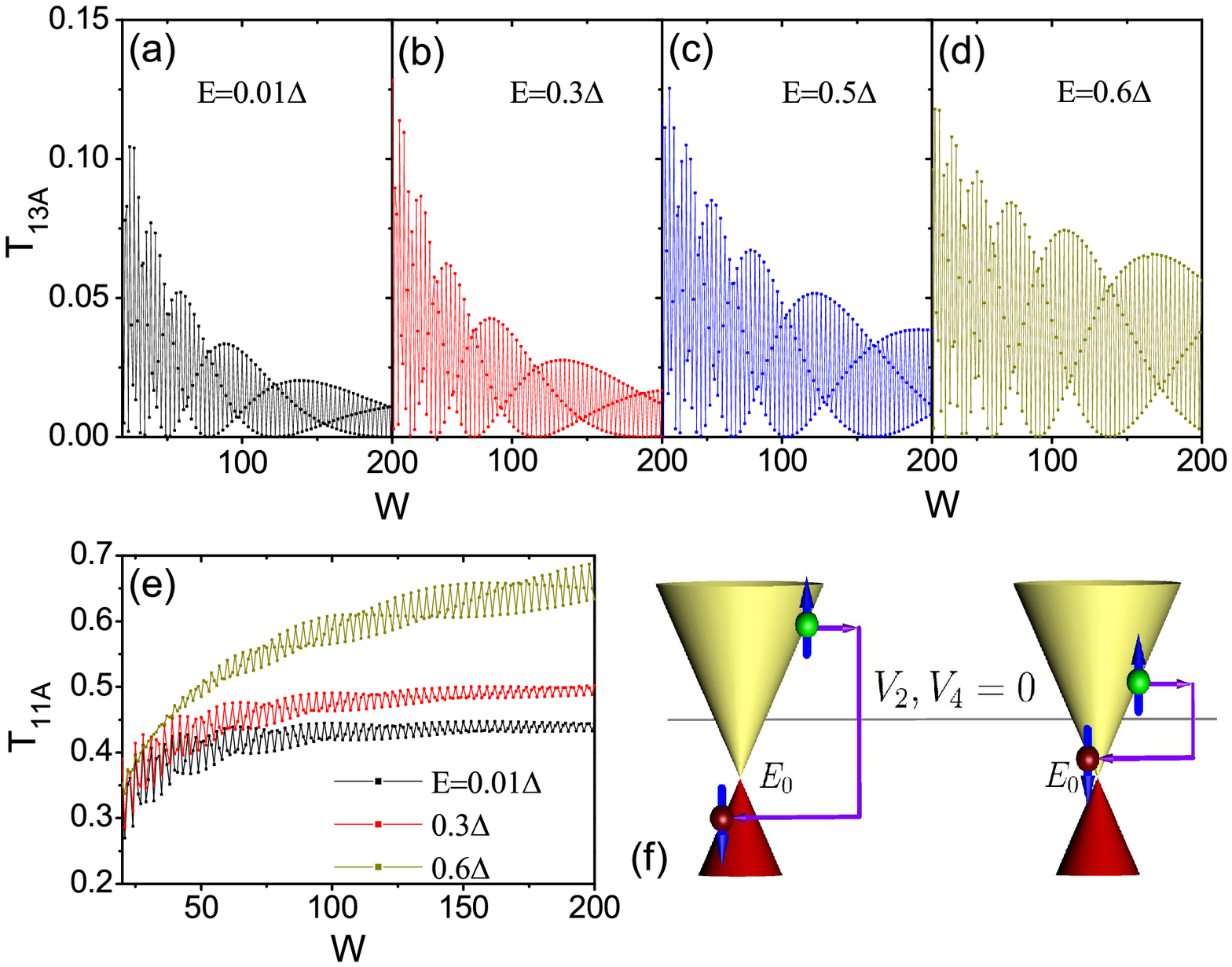}\\
\caption{(color online) (a)-(d) is $T_{13A}$ vs. the width $W$ and
(e) is $T_{11A}$ vs. $W$ with the parameters $E_0=-0.7\Delta$ and
$\theta=0$. (f) is the schematic view of the Andreev retroreflection
and specular Andreev reflection. }
\end{figure}

Next, we focus on the four-terminal device. Fig.2 shows $T_{11A}$
and $T_{13A}$ versus the energy $E$ for two different superconductor
phase differences $\theta\equiv\theta_2-\theta_4=0$ and $\pi$. When
$\theta=0$, $T_{11A}$ and $T_{13A}$ are zero for $|E|>|E_0|$ but
quite large for $|E|<|E_0|$ (see Fig.2a and 2b). This means that at
$\theta=0$ only the retroreflection occurs and the specular
reflection is prohibited. On the other hand, when $\theta=\pi$, the
situation reverses: $T_{11A}$ and $T_{13A}$ are zero for $|E|<|E_0|$
and quite large when $|E|>|E_0|$ (see Fig.2c and 2d). Hence when
$\theta=\pi$, the retroreflection is prohibited and only specular
reflection occurs. Experimentally, the phase difference $\theta$ can
be tuned by varying the super-current between two superconductor
terminals. So the present four-terminal device gives us a handle to
experimentally control and select the Andreev retroreflection and
specular Andreev reflection.

Now we explain why the retroreflection disappears at the phase
difference $\theta=\pi$ while the specular reflection vanishes at
$\theta=0$. In the four-terminal device with two
graphene-superconductor interfaces, two Andreev reflections from
each interface contribute coherently to the resultant Andreev
reflection coefficient. Depending on the phase carried by each
Andreev reflection, the interference can either be constructive or
destructive. For the retroreflection, each reflected hole carries a
phase factor,\cite{ref16} $\theta_{\alpha}$ of the corresponding
superconductor terminal, leading to a total Andreev reflection
coefficient proportional to $|e^{i\theta_2}+e^{i\theta_4}|^2
=|1+e^{i\theta}|^2$, whose value reaches the maximum at $\theta=0$
and minimum at $\theta=\pi$. So the Andreev retroreflection
disappear at $\theta=\pi$ due to the destructive interference.
However, for the specular Andreev reflection, in addition to the
phase difference $\theta$, an extra phase $\pi$ is acquired due to
the reflection between two interfaces when the incident electron and
reflected hole involves different energy bands. The origin of this
extra phase $\pi$ is similar to the $\pi$ junction of the
superconductor-graphene-superconductor device,\cite{ref6} where a
super-current of form $I=I_c \sin(\theta+\pi)$ was found. Due to
this extra phase the total Andreev reflection coefficient is
proportion to $|1+e^{i(\theta+\pi)}|^2$, whose value is zero at
$\theta=0$ resulting a vanishing specular Andreev reflection at
$\theta=0$.

\begin{figure}
\includegraphics[bb=30 31 596 239, scale=0.4, clip=]{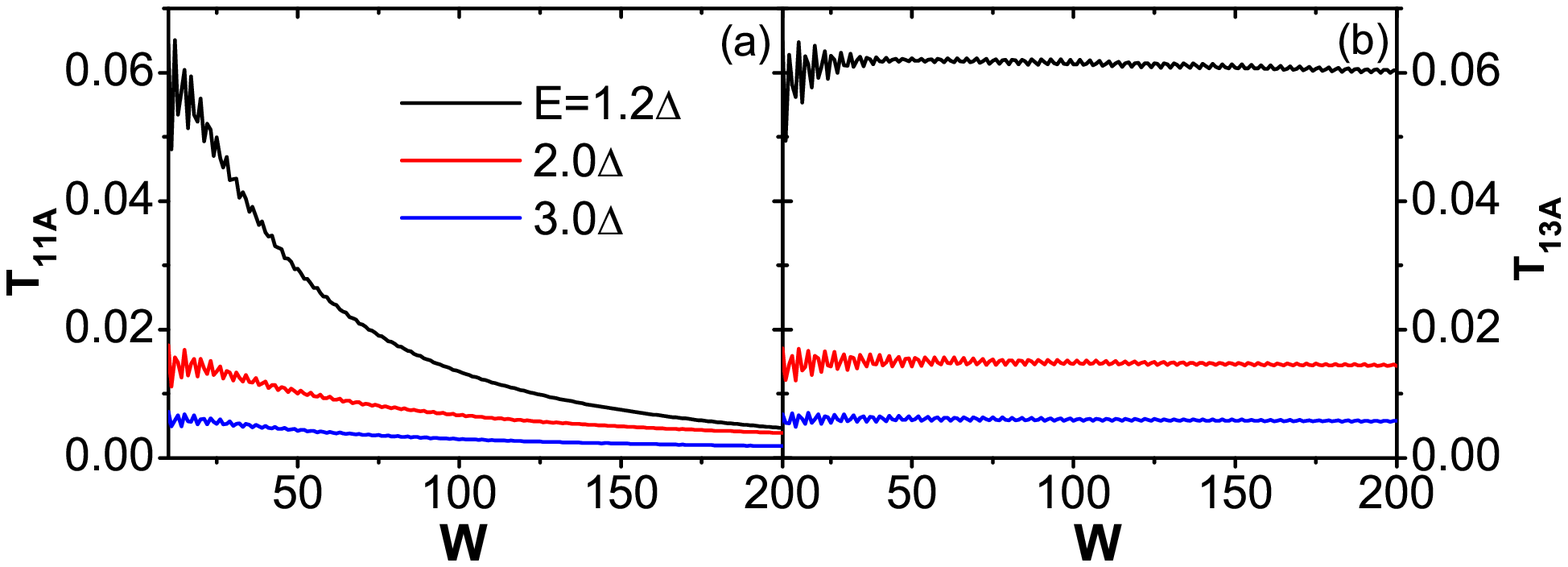}\\
\caption{(color online) $T_{11A}$ and $T_{13A}$ vs. $W$ for
$E_0=-0.7\Delta$ and $\theta=\pi$.}
\end{figure}

Since the reflected hole from the Andreev retroreflection (the
specular Andreev reflection) is along the retro-reflected (the
specular reflected) direction as shown in Fig.1a, Andreev reflection
coefficient $T_{13A}$ ($T_{11A}$) should be zero at $\theta=0$
($\pi$). This does not agree with what we have obtained in Fig.2. We
attribute this phenomenon to the diffraction effect of the reflected
hole in the small device. To verify this statement, we have studied
the size dependence of the Andreev reflection coefficient. Fig.3a-3e
show $T_{13A}$ and $T_{11A}$ versus the width $W$ for $\theta=0$, in
which only the retroreflection occurs. With the increase of the
width $W$, $T_{13A}$ oscillates and decays to zero while $T_{11A}$
increases and saturates at large $W$. This clearly indicates that
the reflected hole exits only from the terminal-1 at large width
$W$. In addition, from the standing wave in the semi-infinite
graphene ribbon, the wavelength of the reflected hole has been
calculated. We found that when the graphene-ribbon width $W$ is in
several tens this wavelength is on the same order of the device
size. So for this width $W$ (several tens), the diffraction effect
is significant. As a result, the reflected hole can exit from both
terminal 1 and 3 leading to non-zero values of both $T_{11A}$ and
$T_{13A}$.

We now examine the size dependence of the specular Andreev
reflection when $\theta=\pi$. We expect that when the width $W$ of
the graphene ribbon increases the reflected hole should go to the
terminal 3 due to the fact that the diffraction effect disappears at
large $W$. Indeed, our numerical result confirms this. From Fig.4 we
see that $T_{11A}$ decays to zero and $T_{13A}$ saturates at large
$W$.

\begin{figure}[htbp]
\begin{center}
\centering
\includegraphics[bb=31 38 614 446, scale=0.38, clip=]{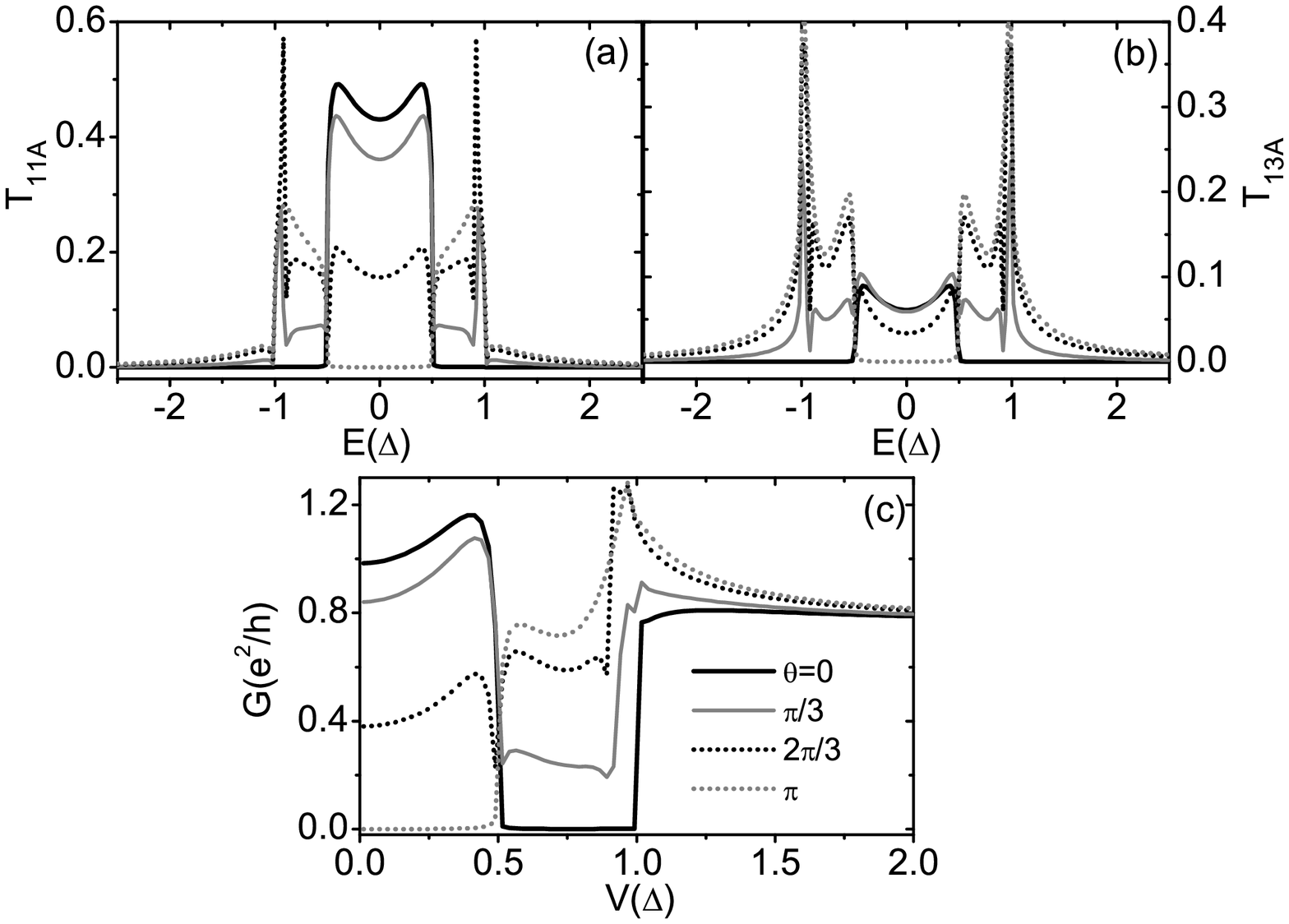}\\
\caption{Upper panel: $T_{11A}$ (a) and $T_{13A}$ (b) vs. $E$ for
different $\theta$. Lower panel: the conductance $G$ of terminal 1
as a function of bias $V$. The parameters are $E_0=-0.5\Delta$ and
$W=50$.}
\end{center}
\end{figure}

From the above discussion, we see that when the wavelength of the
reflected hole is comparable to the width of graphene nanoribbon,
its direction can not be used to distinguish the Andreev
retroreflection and specular Andreev reflection. Nevertheless, these
two kinds of Andreev reflection can manifest their difference in the
four-terminal device by tuning the superconductor phase as
demonstrated above. This also shows that the electron-hole
conversion mechanism, i.e., interband or intraband conversion, is
the fundamental origin for these two kinds of Andreev reflections.

Up to now, we only considered the clean graphene ribbon at
$\theta=0$ or $\pi$. In the presence of the weak impurity
disorder,\cite{ref20} all the results still remain, except that the
boundary of two kinds of Andreev reflection slightly smeared. In
addition, for other $\theta$, both types of Andreev reflection may
occur (see Fig.5a and 5b) due to the incomplete destructive
interference. With the variation of $\theta$ from $0$ to $\pi$, the
specular reflection ($T_{11A}$ and $T_{13A}$ at $|E|>|E_0|$)
gradually increases from zero to the maximum value while the
retroreflection ($T_{11A}$ and $T_{13A}$ at $|E|<|E_0|$) gradually
decreases from the maximum value to zero.

Finally, we investigate the differential conductance $G \equiv
dI_1/dV$ at zero temperature and discuss the experimental
feasibility. By setting the bias $V_1=V_3=V$ and $V_2=V_4=0$, the
direct tunneling $T_{13}$ from the terminal 1 to the terminal 3 does
not contribute to the current. The differential conductance is given
by $G(V)=(2T_{11A}+2T_{13A}+T_{12}+T_{14})2e^2/h$. At small bias
$|eV|<\Delta$, the normal tunneling processes from the terminal 1 to
two superconductor terminal 2, 4 ($T_{12}$ and $T_{14}$) are
forbidden, so the conductance $G(V)$ is directly related to the
Andreev reflection coefficient $T_{11A}+T_{13A}$. Fig.5c shows the
conductance $G$ versus the bias $V$ at different $\theta$. For
$\theta=0$, the specular Andreev reflection vanishes, so $G$ is zero
at $|E_0|<|eV|<\Delta$. On the other hand, for $\theta=\pi$, the
Andreev retroreflection disappears leading to $G=0$ at $|eV|<|E_0|$.
Upon varying $\theta$ from $0$ to $\pi$, the conductance $G$ at
$|eV|<|E_0|$ drops to zero while $G$ at $|E_0|<|eV|<\Delta$
gradually increases from $0$. For the bias $|eV|>\Delta$, $G$ is
always large and weakly depends on the phase difference $\theta$
because of the contribution of the normal tunneling process. Note
that experimentally the graphene nanoribbons\cite{ref17} have been
realized with the superconductor leads attached to graphene
nanoribbons.\cite{ref9} So the proposed device is within the reach
of the present technology and is feasible experimentally.

In conclusion, the interplay of two kinds of Andreev reflections in
a four-terminal graphene-superconductor hybrid device was
investigated. It was found that the Andreev retroreflection and
specular Andreev reflection can be tuned in this system due to
quantum interference. When the superconductor phase difference
$\theta =0$, the specular Andreev reflection is prohibited and only
the Andreev retroreflection occurs. However, for $\theta=\pi$, the
Andreev retroreflection is suppressed and only the specular Andreev
reflection occurs. In addition, in the narrow graphene nanoribbon
with its size comparable to the wavelength of the reflected hole,
the diffraction effect occurs. Then the reflected hole can exit from
both graphene terminals. On the other hand, for large samples, the
diffraction effect disappears and the reflected hole can only
traverse to a particular terminal depending on the kind of Andreev
reflections.

{\bf Acknowledgments:} The work is supported by NSFC under Grant
Nos. 10525418, 10734110, and 10821403 and by 973 Program Project No.
2009CB929103 (Q.F.S.); a RGC grant from the Government of HKSAR
grant number HKU 704308P (J.W.).

\end{document}